\documentclass[10pt,conference]{IEEEtran}
\usepackage{amsfonts}
\usepackage{amssymb,epic,eepic}
\usepackage{amsmath}
\newcommand{\cc}{{\mathbb C}}
\newcommand{\rr}{{\mathbb R}}

\newcommand{\idn}{\mathbf{1}}

\newtheorem{theorem}{Theorem}[section]         
\newtheorem{lemma}[theorem]{Lemma}             
\newtheorem{corollary}[theorem]{Corollary}     

\begin{document}

\title{Structure of Optimal Input Covariance Matrices for MIMO Systems with Covariance Feedback under General Correlated Fading}

\author{Igor Bjelakovi\'c and Holger Boche\\Heinrich-Hertz-Chair for Mobile Communications\\ Berlin University of Technology\\Werner-von-Siemens-Bau (HFT 6), Einsteinufer 25, 10587 Berlin, Germany\\Email: \{igor.bjelakovic, holger.boche\}@mk.tu-berlin.de}

\maketitle
\begin{abstract}
We describe the structure of optimal Input covariance matrices for single user multiple-input/multiple-output (MIMO) communication system with covariance feedback and for general correlated fading. Our approach is based on the novel concept of right commutant and recovers previously derived results for the Kronecker product models. Conditions are derived which allow a significant simplification of the optimization problem.
\end{abstract}
\section{Introduction}
Since the seminal work of Telatar \cite{telatar} on the Shannon capacity of multi-antenna wireless systems, this area has attracted a lot of attention. The deveploment has started with the investigation  of the capacity of single-user MIMO systems. Many results on the capacity for different types of channel state information at the transmitter and/or receiver are known. The achieved progress in this field was the key element, that MIMO systems are already used in existing systems. One important research topic on MIMO systems is the impact of correlation of the channel matrix on the achievable capacity \cite{foschini}-\cite{eduard-holger}. A lot of results are known in this area, but most of the works are using the assumption, that the channel covariance matrix is the Kronecker product of the covariance matrices of the transmit and receive antennas \cite{chuah, shiu}. In the following paper the general case is analyzed. \\
The paper is organized as follows: In Section II we review shortly the model and formulate the main problem. Additionally we divide the set of variance matrices into two classes of separable and entangled positive semidefinite matrices, a definition borrowed from quantum information theory. This separation shall help us to present our results for the class of the separable matrices which is easier to deal with, followed by an extension of results to entangled matrices. Section III starts with a novel concept of right commutant which is the key ingredient in our approach. It can be seen as a characterization of one-sided invariant subspaces for the given channel variance matrix (cf. Lemma \ref{trivial-lemma}) or, alternatively, as description of symmetries of the channel variance matrix (cf. Lemma \ref{trivial-lemma-2}.1). Our subsequent results in Section III rely hardly on that concept, which, combined in a appropriate way with some simple concavity considerations \footnote{After finishing this paper we learned that Tulino, Lozano and Verd\'u \cite{verdu} used the concavity of the capacity in a similar way to characterize optimal covariances for channels with independent columns and symmetric joint distribution.}, turns out to be rather powerful tool. For example, we do not need any majorization results/considerations which are the basis of results in \cite{goldsmith, eduard-holger}. Our main result, Theorem \ref{main-th-general}, is a characterization of optimal input variance matrices.\\
\emph{Notation and Preliminaries}
We shall denote matrices by capital letters, e.g. $H$. The hermitian conjugate (adjoint) is denoted by $(\cdot)^{H}$ while $(\cdot)^{t}$ is reserved for the transpose of a matrix. The set of $N\times N$ matrices with complex entries is abbreviated by $\mathbb{M}(N,\cc)$ and $A\otimes B$ denotes the tensor product (Kronecker product) of matrices $A$ and $B$. $\idn_{N}$ is the $N\times N$ unity matrix. $\textrm{diag}(Q_{1},\ldots ,Q_{c})$ is the shorthand for the matrix which has the matrices $Q_{1},\ldots ,Q_{c}$ as its diagonal entries and $0$s else, the size of the diagonal blocks will be specified in each particular case. $\textrm{tr}(A)$ is the trace of the matrix $A$ and $H\sim \mathcal{N}(0,\Sigma)$ means that the complex valued random matrix $H$ of prescribed size is normally distributed with mean $0$ and variance $\Sigma$.  \\
We shall introduce some simple concepts from the theory of $*$-algebras of matrices which will be helpful in this paper (cf. \cite{takesaki} chap. I for more information). A $*$-\emph{algebra} $\mathcal{A}$ in $\mathbb{M}(N,\cc)$ is a linear subspace which is closed under matrix multiplication and under the action of $(\cdot )^{H}$-operation. It can be shown \cite{takesaki} that each $*$-algebra of matrices has a multiplicative unit. $*$-algebras appearing in this paper shall have $\idn_N$ as the unit element with respect to the matrix multiplication. A (orthogonal) projection $P\neq 0$ is called \emph{minimal projection} in $\mathcal{A}$ if $P\in \mathcal{A}$ and $Q\le P$ for any projection $ Q\in \mathcal{A}$ implies $Q=0$ or $P=Q$. Equivalently, a non-zero projection $P\in \mathcal{A}$ is minimal if and only if $P\mathcal{A} P=\cc P$.
By a \emph{resolution of identity} in $\mathcal{A} $ we mean a set of mutually orthogonal projections $\{P_i\}_{i=1}^{c}\subset \mathcal{A}$ that satisfies $\sum_{i=1}^{c}P_i =\idn$, where $\idn$ denotes the multiplicative unit in $\mathcal{A}$.\\
If $A\in\mathcal{A}\subset \mathbb{M}(N,\cc)$ is hermitian or normal matrix, then we can represent it according to the spectral theorem as $A=\sum_{\lambda \in \sigma(A)} \lambda P_{\lambda}$,
where $\sigma(A)$ denotes the spectrum (set of eigenvalues) and $P_{\lambda}$ is the projection onto the eigenspace corresponding to the eigenvalue $\lambda$. By defining properties of a $*$-algebra, with $A\in \mathcal{A}$ we also have $g(A)\in \mathcal{A}$ for each complex valued polynomial. It is easily seen that for each $\lambda\in\sigma(A)$ there is complex valued polynomial $g_{\lambda}$ with $g_{\lambda}(A)=P_{\lambda}$ and hence $P_{\lambda}\in \mathcal{A}$ for all $\lambda\in \sigma(A)$, a fact which will be useful in the proof of Lemma \ref{trivial-lemma} below.\\
Finally, we recall a way of viewing a tensor product of matrices as a linear map which will be necessary in the last part of the paper: For $A\in \mathbb{M}(M,\cc), B\in \mathbb{M}(N,\cc)$ we consider the tensor product $A\otimes B$ and an $M\times N$ matrix $H$. Then it is easily seen using rank one $M\times N $ matrices that the canonical action of $A\otimes B$ on $H$ is given by $(A\otimes B)(H)=AHB^{t}$. This action extends to arbitrary elements of $\mathbb{M}(M,\cc)\otimes \mathbb{M}(N,\cc)$ by linearity, since each $\Sigma \in \mathbb{M}(M,\cc)\otimes \mathbb{M}(N,\cc)$ can be written as a complex linear combination of such elementary tensors $A\otimes B$.  
\section{Model and Problem Formulation}
We focus on a single point-to-point wireless communication system using $N$ transmit and $M$ receive antennas. We assume, that the behavior of the channel can be described by the well known narrow-band flat fading channel model, i.e.
\begin{displaymath}
y=H x +n,
\end{displaymath}
where $x$ is the $N$ dimensional transmit vector, $y$ is the $M$ dimensional receive vector, $H$ is the $M \times N$ channel matrix, and the $M$ components $n_k$ of the noise vector $n$ are assumed to be i.i.d. complex circularly symmetric Gaussian distributed with mean 0 and variance $\sigma_n^2$.
For the channel matrix $H$ we will use a more general correlation model than \cite{goldsmith, eduard-holger} to present our ideas in the most transparent way which allows a direct comparison with the existing results. Then we shall show that this correlation model already incloses the full complexity of the general case. 
The channel matrix in this special case can be described as follows:
\begin{eqnarray}\label{channel-model}
H=\sum_{i=1}^{s}R_{i}^{\frac{1}{2}}W_{i}T_{i}^{t\frac{1}{2}},
\end{eqnarray}
where $W_{i}$ are i.i.d. zero mean , mutually independent complex Gaussian $M\times N$ matrices and the positive semidefinite $M\times M$ resp. $N\times N$ matrices $R_{i}$ resp. $ T_{i}$ are related to the variance $\Sigma$ of $H$ by 
\begin{eqnarray}\label{separable-variance}
\Sigma =\sum_{i=1}^{s}R_{i}\otimes T_{i}, 
\end{eqnarray}
where $\Sigma:=\mathcal{E}(H\otimes \overline{H})$ which has components $\mathcal{E}(H_{i,j}\overline{H_{l,m}})$.
Observe that, since we are dealing with \emph{complex} matrices, $A\ge 0$ implies that $A$ is hermitian.\\
\emph{Remark:} Note that such decompositions into a sum of tensor products of positive semidefinite (PSD) matrices are, in general, non-unique: a simple example is given in the symmetric case of two transmit and two receive antennas with the variance matrix $\Sigma=\idn\otimes\idn$ which can be alternatively decomposed into $\Sigma =\sum_{i=1}^{2}\idn \otimes e_{i}e_{i}^{H}$, $\{e_{1},e_{2}\}$ being any orthonormal basis in $\cc^{2}$. This non-uniqueness with respect to decompositions corresponds to the freedom of choice in the particular realization of random variables distributed according to a given probability distribution.\\
PSD matrices acting on $\cc^{M}\otimes\cc^{N}$ that allow a decomposition as in (\ref{separable-variance}) with PSD summands are called \emph{separable} in quantum information theory. Otherwise we say that they are \emph{entangled} (cf. \cite{werner, gurvits} and references therein). The simplest example of an entangled PSD matrix is given by $gg^{H}$, where $g:=e_{1}\otimes e_{1}+e_{2}\otimes e_{2}$ and $\{e_{1},e_{2}\}$ being canonical basis of $\cc^{2}$.\\
A handy sufficient criterion for separability of a given PSD matrix over $\cc^{M}\otimes\cc^{N}$ is given in \cite{gurvits}: 
\begin{theorem}[Gurvits/Barnum]\label{gurvits-barnum}
A PSD matrix $\Sigma$ is separable if $||\Sigma-\idn_{M}\otimes\idn_{N}||_{2}\le 1$, where $||\cdot ||_{2}$ denotes the Hilbert-Schmidt norm on matrices (i.e. $||A||_{2}:=\sqrt{(A,A)_{HS}}:=\sqrt{\textrm{tr}(A^{H}A)}$).
\end{theorem}
In the following paper we assume, that the receiver knows the channel perfectly, and the transmitter has only knowledge of the channel covariance matrix $\Sigma$. As a consequence, the channel state information at the transmitter is a  deterministic function of the channel state information at the receiver. Under this condition the ergodic capacity of the considered MIMO system is given by
\begin{eqnarray}\label{channel-state-information}
C=\underset{\substack{tr(Q)\le p \\ Q \ge 0}}{\max}\mathcal{E}(\log \det (\idn_M +\frac{1}{\sigma_n^2}H Q H^H)),
\end{eqnarray}
as it is easily seen using the results of \cite{caire}.
The optimization problem ($\ref{channel-state-information}$) is a convex smooth optimization problem. The capacity $C=C(\hat{Q})$ for an optimal transmit covariance matrix $\hat{Q}$ is achieved by transmitting independent complex circular Gaussian symbols along the eigenvectors of $\hat{Q}$, and the powers are allocated according to the eigenvalues of the matrix $\hat{Q}$ \cite{simon}-\cite{eduard-holger}.
\section{Results}
 For a given variance matrix $\Sigma \in \mathbb{M}(M,\cc)\otimes \mathbb{M}(N,\cc)$ we define the ``right'' commutant
\[\mathcal{C}_{\Sigma}:=\{ A\in \mathbb{M}(N,\cc)| (\idn_{M}\otimes A)\Sigma =\Sigma (\idn_{M}\otimes A)\},\]
and consider any resolution of unity consisting of mutually orthogonal minimal projections in $\mathcal{C}_{\Sigma} $, i.e. $\idn_{N}=\sum_{i=1}^{t}P_{i}$ with $P_{i}\in \mathcal{C}_{\Sigma} $ minimal and $P_{i}P_{j}=\delta_{i j}P_{i}$.\\
\emph{Example 1.} If the variance matrix is given by $\Sigma =R\otimes T$ then we have $ \mathcal{C}_{\Sigma}=\{A\in \mathbb{M}(N,\cc)| AT=TA\}$,
and each set of mutually orthogonal minimal projections in $\mathcal{C}_{\Sigma}$ adding to $\idn_{N}$ is given by projections onto the one-dimensional subspaces spanned by the eigenvectors of $T$.\\
Some simple observations concerning the concept of right commutant are collected for ease in the following
\begin{lemma}\label{trivial-lemma}
Let $\Sigma$ be a PSD matrix in $ \mathbb{M}(M,\cc)\otimes \mathbb{M}(N,\cc) $ then we have:
\begin{itemize}
\item[1.] $\mathcal{C}_{\Sigma} $ is a subalgebra of $ \mathbb{M}(N,\cc) $ containing $\idn_N$ which is closed under $(\cdot)^{H}-$operation, i.e. $\mathcal{C}_{\Sigma}$ is a $*$-algebra.
\item[2.] Let $\{P_{i}\}_{i=1}^{u}$ and $\{Q_{j}\}_{j=1}^{v}$ be resolutions of identity consisting of minimal projections in $\mathcal{C}_{\Sigma} $ . Then $u=v$ and there is a permutation $\pi$ of $\{1,\ldots ,u\}$ such that $\textrm{tr}(P_{i})=\textrm{tr}(Q_{\pi (i)})$ for all $i\in \{1,\ldots ,u\}$.
\item[3.] If $\Sigma$ is separable and if $\{P_{j}\}_{j=1}^{u} $ is a resolution of identity consisting of minimal projections in $\mathcal{C}_{\Sigma} $, then there is a decomposition of $\Sigma$ into sum of tensor products of PSD matrices
\[\Sigma =\sum_{i=1}^{s}R_{i}\otimes T_{i},\]
satisfying $T_{i}P_{j}=P_{j}T_{i}$ for all $i\in\{1,\ldots ,s\}$ and $j\in\{1,\ldots ,u\}$.
\end{itemize}
\end{lemma}  
\emph{Remark:} Our right commutant $\mathcal{C}_{\Sigma}$ is a close relative of the concept of commutant which is widely used in the theory of operator algebras and quantum information theory. And, indeed, the proof of the properties stated in Lemma \ref{trivial-lemma} consist of some standard conclusions, at least for those already familiar with the usual commutant from the theory of operator algebras. For the ease of reading we include this short proof.   \\
\emph{Proof of Lemma \ref{trivial-lemma}:} The first item is easily checked by inspection and is standard in the theory of matrix (operator) algebras (cf. \cite{takesaki}). For the second item, note that each $P_{i}Q_{j}P_{i}$ is hermitian and contained in $\mathcal{C}_{\Sigma} $. It is well known that then all spectral projections of $P_{i}Q_{j}P_{i}$ are also contained in $\mathcal{C}_{\Sigma} $. Using this fact it is easy to deduce a contradiction to the assumed minimality of the involved projections unless $u=v$. The second part is then easily obtained.\\
The third item follows from the relation 
\[\Sigma = \sum_{j=1}^{u}(\idn_{M}\otimes P_{j})\Sigma (\idn_{M}\otimes P_{j}),\] combined with $\Sigma =\sum_{l=1}^{n} \tilde{R}_{l}\otimes \tilde{T}_{l}$ where $\tilde{R}_{l} $ and $\tilde{T}_{l} $ are PSD, which is ensured by separability of $\Sigma$. Indeed, we merely have to set 
\[ R_{i}:=\tilde{R}_{i}\quad \textrm{and}\quad T_{i}:=\sum_{j=1}^{u}P_{j}\tilde{T}_{i}P_{j},\]
and we arrive at the desired conclusion of the lemma. $\qquad \Box$ \\
\emph{Remark:} As we will show in the following the minimal projections $\{P_{j}^{t}\}_{j=1}^{u}$ shall serve as the starting point of block-diagonalization procedure for optimal input covariance matrices. The second part of Lemma \ref{trivial-lemma} ensures that no particularly chosen minimal resolution of identity is preferred, i.e. the dimensions of the corresponding ranges of considered projections are equal up to a permutation. 

Unfortunately, there are cases where the algebra $\mathcal{C}_{\Sigma} $ is trivial, i.e. consists of complex multiples of $\idn_{N}$ as the following example shows:\\
\emph{Example 2.} Let $M=2=N$ and $\Sigma=e_{1}e_{1}^{H}\otimes e_{1}e_{1}^{H}  + e_{2}e_{2}^{H}\otimes g g^{H}$ ,
where $\{e_{1},e_{2}\}$ denotes the canonical basis in $\cc^{2}$ and $g=\frac{1}{\sqrt{2}}(e_{1}+e_{2})$. Let $P\in \mathcal{C}_{\Sigma} $ be a projection, then we have $(\idn_{M}\otimes P)\Sigma =\Sigma (\idn_{M}\otimes P)$. Inserting this into the expression for $\Sigma$ above and multiplying with $e_{i}e_{i}^{H}\otimes\idn_{N}$ for $i=1,2$ we end up with two equations $ e_{1}e_{1}^{H}P=P e_{1}e_{1}^{H} \textrm{ and } g g^{H} P=P g g^{H}$.
A simple calculation shows that $P=\alpha \idn_{N}$ with $\alpha\in \rr_{+}$ and hence $P=0$ or $P=\idn_{N}$.\\
In the following we separate our presentation in two parts; in the first we consider the separable variance matrices while in the second no restrictions on channel matrices $H$ are assumed. This separation, although not necessary from the viewpoint of mathematics, has the advantage that we can first present our ideas in a situation which is close in the spirit to the previous work of Jafar/Wishwanath/Goldsmith \cite{jafar,goldsmith} and Jorswieck/Boche \cite{eduard-holger}, and then we show that the result extends immediately to the general case.
\subsection{Optimal Input Covariance Matrices: Separable Case}
Now, we can describe the optimal input matrix in the case where $\Sigma$ is separable and $\mathcal{C}_{\Sigma}$ contains non-trivial minimal projections, i.e. not equal $\idn_{N}$.\\
Choose any resolution of identity consisting of minimal mutually orthogonal projections $\mathcal{C}_{\Sigma}^{t} $ (the transpose of $\mathcal{C}_{\Sigma} $) , denoted by $\{P_{j}\}_{j=1}^{c}$, and a decomposition of $\Sigma$ with properties given in Lemma \ref{trivial-lemma}.3 with respect to $\{P_{j}^{t}\}_{j=1}^{c}$, a resolution of identity  consisting of minimal projections in $\mathcal{C}_{\Sigma}$. Then there is a unitary $U$ such that $T_{i}^{t}= U\textrm{diag}(T_{i}(1),\ldots, T_{i}(c))U^{H}$ for all $i\in\{1,\ldots ,s\}$, where the matrices $T_{i}(j)$ map the range of $P_{j}$ into itself, i.e. each $T_{i}^{t}$ is block-diagonal in the basis given by the unitary matrix $U$. 
\begin{theorem}\label{main-th-separable}
Suppose that the variance matrix $\Sigma$ of $H\sim\mathcal{N}(0,\Sigma)$ is separable and that $\mathcal{C}_{\Sigma}\neq \cc \cdot \idn_{N}  $. Then the capacity achieving covariance matrix $Q$ can be chosen such that
\[Q=U\textrm{diag}(Q_{1},\ldots ,Q_{c})U^{H},\]
where each $Q_{j}$ maps the range of $P_{j}$ into itself, $j\in\{1,\ldots ,c\}$. 
\end{theorem}
\emph{Proof:} Suppose that we are given any capacity achieving covariance matrix $Q$, i.e.
\[ C=C(Q)=\mathcal{E}\left(\log\det \left(\idn_{M}+\frac{HQH^{H}}{\sigma_{n}^{2}}\right)\right).\]
Due to our system assumption, the last expression is written as
\[C=\mathcal{E}(\log\det (\idn_{M}+\frac{\sum_{i=1}^{s}R_{i}^{\frac{1}{2}}W_{i}T_{i}^{t\frac{1}{2}} Q \sum_{l=1}^{s}T_{l}^{t \frac{1}{2}}W_{l}^{H}R_{l}^{\frac{1}{2}} }{\sigma_{n}^{2}})).\]
Now, we insert the relation 
\[T_{i}^{t}=U\textrm{diag}(T_{i}(1),\ldots, T_{i}(c))U^{H}=:U\tilde{T}_{i}U^{H},\]
 with $\tilde{Q}:=U^{H}QU$ fulfilling $\textrm{tr}(Q)=\textrm{tr}(\tilde{Q})$ and arrive at
\begin{eqnarray}\label{c-tilde}
C &=&\mathcal{E}(\log\det (\idn_{M}+\frac{\sum_{i,l=1}^{s}R_{i}^{\frac{1}{2}}W_{i}\tilde{T}_{i}^{\frac{1}{2}} \tilde{Q} \tilde{T}_{l}^{ \frac{1}{2}}W_{l}^{H}R_{l}^{\frac{1}{2}} }{\sigma_{n}^{2}})) \nonumber \\
&=:& \tilde{C}(\tilde{Q})
\end{eqnarray}
where we have used that the random matrices $W_{i}$ and $W_{i}U$ have the same probability distribution since each $W_{i}$ is i.i.d. Gaussian and the $W_{i}$'s are jointly independent. The transformed matrix $\tilde{Q}$ can be written as a block matrix with respect to the transformation $U$ induced by the set $\{P_{j}\}_{j=1}^{c} $ of minimal projections in $\mathcal{C}_{\Sigma}^{t}$:
\[\tilde{Q} =\left(
\begin{array}{ccc}
Q_{11} & Q_{12} & \ldots  Q_{1c} \\
Q_{21} & Q_{22} & \ldots   Q_{2c} \\
\vdots & \vdots & \ddots   \\
Q_{c1} & Q_{c2} & \ldots  Q_{cc}
\end{array}
\right) 
. \]
We consider the \emph{unitary and hermitian} matrix 
\[U_{1}:= \textrm{diag}(\idn_{P_{1}}, -\idn_{P_{2}}, -\idn_{P_{3}},\ldots ,-\idn_{P_{c}}),\]
where $\idn_{P_{j}}$ denotes the matrix acting as the identity on the range of $P_{j}$. Then we have $U_{1}\tilde{T}_{i}U_{1}=\tilde{T}_{i}$,
\[ \tilde{Q}_{1}:=\frac{1}{2}(\tilde{Q}+U_{1}\tilde{Q}U_{1})= \left( \begin{array}{ccc}
Q_{11} & 0 & \ldots  0 \\
0 & Q_{22} & \ldots   Q_{2c} \\
\vdots & \vdots & \ddots   \\
0 & Q_{c2} & \ldots  Q_{cc}
\end{array}
\right),\]
and $\textrm{tr}(\tilde{Q})=\textrm{tr}(\tilde{Q}_1)$.\\
Due to the concavity of the functional $\tilde{C}$ defined by the last eqn. in (\ref{c-tilde}) we end up with
\begin{eqnarray}
C &\ge & \tilde{C}(\tilde{Q}_{1})\ge \frac{1}{2}\tilde{C}(\tilde{Q})+\frac{1}{2}\tilde{C}(U_{1}\tilde{Q}U_{1})=\tilde{C}(\tilde{Q})\nonumber \\
&= & C,
\end{eqnarray}
where we have used $U_{1}\tilde{T}_{i}U_{1}=\tilde{T}_{i}$ in the first equality.
In the next step we consider the unitary and hermitian matrix $U_{2}$ given by
\[U_{2}:= \textrm{diag}(\idn_{P_{1}}, \idn_{P_{2}}, -\idn_{P_{3}},\ldots ,-\idn_{P_{c}}),\] and can define in a similar way a matrix $\tilde{Q}_{2}:=\frac{1}{2}(\tilde{Q_{1}}+U_{2}\tilde{Q_{1}}U_{2}) $ and show analogously that
$\tilde{C}(\tilde{Q}_{2})=C$ holds. Continuing this procedure we arrive at the claimed conclusion of the theorem.$\qquad\qquad\Box$ \\
Note that, as mentioned previously, in the case $\Sigma=R \otimes T$ the resolution of identity $\{P_{j}\}_{1}^{c}$ consists of one-dimensional projections, i.e. $c=N$ and we recover the results of \cite{goldsmith, eduard-holger} that the optimal transmission strategy consists of sending independent circularly symmetric gaussian inputs along the eigenvectors of $T$.  

\subsection{Optimal Input Covariance Matrices: General Case}
If we examine carefully our construction in the proof of theorem \ref{main-th-separable} we see that we have needed only the concavity of the capacity functional together with the fact that $U_{j}\tilde{T}_{i}U_{j}=\tilde{T}_{i} $ which means that applying $U_{j}$ does not change the probability distribution of the considered random matrix $H$. Hence, in order to extend our proof to the case of general random matrices $H\sim \mathcal{N}(0,\Sigma)$ we merely have to consider the basis-free versions of hermitian and unitary matrices $U_{j}=2(P_{1}+\ldots + P_{j})-\idn_{N}, j=1,\ldots ,c $ which realize our block-diagonalization. Taking into account the first part of Lemma \ref{trivial-lemma-2} below, that contains the description of the symmetries of the channel at our disposal, we conclude that Theorem \ref{main-th-separable} extends \emph{mutatis mutandis} to the general situation. The only change is that we drop the condition of separability we have supposed in the statement of Theorem \ref{main-th-separable}:
\begin{theorem}\label{main-th-general}
Let $H\sim \mathcal{N}(0,\Sigma) $ be a random $M\times N$ channel matrix and suppose that $\mathcal{C}_{\Sigma}\neq \cc \idn_N$. Then the capacity achieving covariance matrix $Q$ can be chosen such that
\[Q=U\textrm{diag}(Q_{1},\ldots ,Q_{c})U^{H},\]
where $Q_{j}$ maps the range of $P_{j}$ into itself, $\{P_{j}\}_{j=1}^{c}$ denotes any resolution of identity consisting of minimal projections in $\mathcal{C}_{\Sigma}^{t}$ and $U$ is any unitary matrix which diagonalizes all $P_{j}$ simultaneously.
\end{theorem}

We now use Theorem \ref{main-th-general} for a further analysis of our optimization problem. We use the structure
\begin{displaymath}
Q=[U^1, \ldots, U^c]\textrm{diag}(Q_1, \ldots, Q_c)[U^1, \ldots, U^c]^H
\end{displaymath}
of the optimal transmit covariance matrix Q. The block $Q_i$ has the dimension $l_i \times l_i$ and the corresponding unitary matrix $U^i$ has the size $M \times l_i$. We have $\sum \limits_{i=1}^c l_i=N$. If we use the matrix $H_i=H U^i$, then we have for the optimal transmit covariance matrix
\begin{displaymath}
C=I(Q)=\mathcal{E}(\log \det (\idn_{M} +\frac{1}{\sigma_n^2}\sum\limits_{l=1}^c H_{l}Q_{l}H_l^H)).
\end{displaymath}
Thus the optimal block matrix $\textrm{diag}(Q_1, \ldots, Q_c)$ can be calculated as the solution of
\begin{displaymath}
\underset{\substack{Q_{l}\ge 0 \\ \sum\limits_{l=1}^c tr(Q_l) \le p}}{\max} \mathcal{E}(\log \det (\idn_{M} +\frac{1}{\sigma_n^2}\sum\limits_{l=1}^c H_{l}Q_{l}H_l^H)).
\end{displaymath}

As a consequence of this simple observation and Theorem $\ref{main-th-general}$ we achieve the following corollary.

\begin{corollary}\label{Cor1}
The block matrix $\textrm{diag}(\hat{Q}_1, \ldots, \hat{Q}_c)$ is the optimal block matrix if and only if, there exists a $\mu > 0$ and positive semidefinite matrices $\Psi_1, \ldots, \Psi_c$, such that $\hat{Q}_k\ge 0, 1\le k\le c$, 

\begin{equation*}
\begin{split}
\frac{1}{\sigma_n^2} \mathcal{E}(\textrm{tr}(H_k^H(\idn_{M} +\sum\limits_{l=1}^c H_l\hat{Q}_{l}H_l^H)^{-1}H_k)=\mu \idn_{l_{k}} -\Psi_k, \\ \textrm{tr}(\Psi_k\hat{Q}_k)=0, \quad 1\le k\le c, 
\end{split}
\end{equation*}
and
\begin{displaymath}
\sum\limits_{l=1}^c \textrm{tr}(\hat{Q}_l)=p
\end{displaymath}
holds.
\end{corollary}

\emph{Remark:} For the classical correlation scenario $\sum = R\otimes T$ we have again $c=N, l_1=\ldots =l_N=1$, and $\hat{Q}=\textrm{diag} (\hat{p}_1, \ldots, \hat{p}_N), \hat{p}_l\ge 0$, where the $\hat{p}_l$ are the solution of the well known power optimization problem \cite{goldsmith, eduard-holger}.\\
The following Lemma \ref{trivial-lemma-2} gives a further description of the optimal transmit covariance matrices.

\begin{lemma}\label{trivial-lemma-2}
 Consider any $M\times N$ random channel matrix $H\sim \mathcal{N}(0,\Sigma)$ and let $U$ be a unitary $N\times N$ matrix. Then:
\begin{itemize}
\item[1.] The channel matrices $H$ and $HU$ have equal probability density functions iff $U^{t}\in \mathcal{C}_{\Sigma}$, or equivalently $U\in  \mathcal{C}_{\Sigma}^{t}$. 
\item[2.] If $Q^{(1)}$ and $Q^{(2)}$ are capacity achieving PSD matrices, i.e. $C(Q^{(1)})=C(Q^{(2)})$, with $\textrm{tr}(Q^{(1)})=p=\textrm{tr}(Q^{(2)})$ then
\begin{eqnarray}\label{q1-q2} 
HQ^{(1)}H^{H}=HQ^{(2)}H^{H}\quad \textrm{a.s.},
\end{eqnarray}
with respect to the law of $H$. 
\end{itemize}
\end{lemma}
\emph{Proof:} 1. The first statement is easily obtained by using change of variables. For reader's convenience we give some crucial steps: First, the variances $\Sigma$ of $H$ resp. $\Sigma_{U}$ of $HU$ are related by $\Sigma_{U}=(\idn_{M}\otimes U^{t H})\Sigma (\idn_{M}\otimes U^{t}) $. This can be easily verified using change of variables formula and observing that each tensor product $A\otimes B\in \mathbb{M}(M,\cc)\otimes \mathbb{M}(N,\cc)$ canonically induces a linear map on $M\times N$ matrices by assignment $H\mapsto AHB^{t}$. Note that the probability density function of the channel matrix can be written as
\[f(H)=K e^{-\scriptsize{\frac{1}{2}}(H,\Sigma^{\scriptsize{-1}}H)_{HS}},\]
where $(\cdot , \cdot )_{HS}$ denotes the Hilbert-Schmidt inner product and $K$ is the normalization constant.
The conclusion of the first part of the lemma is now obvious.   \\
2. According to our assumption and due to the concavity of the capacity functional we may conclude that
\[ C=C(\scriptsize{\frac{1}{2}}Q^{(1)} +\scriptsize{\frac{1}{2}}Q^{(2)})=\frac{1}{2}C(Q^{(1)})+\frac{1}{2}C(Q^{(2)}).\]
Moreover, since the functional $\log\det (\cdot )$ is concave we see that for $\tilde{Q}=\frac{1}{2}(Q^{(1)}+Q^{(2)})$
\begin{eqnarray*}
\begin{split}
\log\det  \left(\idn_{M} +  \frac{H\tilde{Q}H^{H}}{\sigma_{n}^{2}}\right)   
= \frac{1}{2}\log\det\left(\idn_{M}+\frac{HQ^{(1)}H^{H}}{\sigma_{n}^{2}}\right)\\
+ 
\frac{1}{2}\log\det\left(\idn_{M}+\frac{HQ^{(2)}H^{H}}{\sigma_{n}^{2}}\right)
\end{split}
\end{eqnarray*}
holds almost surely with respect to the probability distribution of the channel matrix $H$. This last equation, in turn, is equivalent to
\begin{eqnarray}\label{det-log-conc}
\det  \left(\idn_{M}  +\frac{ H\tilde{Q}H^{H}}{\sigma_{n}^{2}}\right)&=& \det\left(\idn_{M}+\frac{HQ^{(1)}H^{H}}{\sigma_{n}^{2}}\right)^{\scriptsize{\frac{1}{2}}}\nonumber \\ 
&\times & \det\left(\idn_{M}+\frac{HQ^{(2)}H^{H}}{\sigma_{n}^{2}}\right)^{\scriptsize{\frac{1}{2}}}
\end{eqnarray}
almost surely. Now, recall the Minkowski's determinant inequality and the $\log$-concavity of the determinant (cf. \cite{horn-johnson}) which can be stated as the following chain of inequalities:
\begin{eqnarray}\label{minkowski}
\det(\scriptsize{\lambda} A +(1-\scriptsize{\lambda})B)&\ge & \!\! (\lambda \det (A)^{\scriptsize{\frac{1}{M}}} \nonumber \\
&+ & (1-\lambda)\det (B)^{\scriptsize{\frac{1}{M}}})^{\scriptsize{M}}\nonumber \\
&\ge & \det (A)^{\lambda}\det (B)^{1-\lambda},
\end{eqnarray}
for $\lambda \in (0,1)$ and $A,B\in \mathbb{M}(M,\cc)$ positive definite. The equality appears in the first inequality iff $A=\alpha B$ with $\alpha >0$, while the equality in the second line is obtained iff $\det (A)=\det (B)$. Hence the overall equality in (\ref{minkowski}) can appear iff $A=B$. Translating this to our eqn. (\ref{det-log-conc}) we see that
\[ \idn_{M}+\frac{ HQ^{(1)}H^{H}}{\sigma_{n}^{2}}=\alpha (H)\left(\idn_{M}+\frac{HQ^{(2)}H^{H}}{\sigma_{n}^{2}}\right),\]
a.s. with a measurable function $\alpha$ which is almost surely positive and
\[ \det\left(\idn_{M}+\frac{HQ^{(1)}H^{H}}{\sigma_{n}^{2}}\right)= \det\left(\idn_{M}+\frac{HQ^{(2)}H^{H}}{\sigma_{n}^{2}}\right)\quad \textrm{a.s.}\]
These two relations lead immediately to
\begin{eqnarray*}
  \idn_{M}+\frac{HQ^{(1)}H^{H}}{\sigma_{n}^{2}}=\idn_{M}+\frac{HQ^{(2)}H^{H}}{\sigma_{n}^{2}} \quad \textrm{a.s.}\quad\quad \Box
\end{eqnarray*}
\emph{Remark:} As the proof shows, the second part of our Lemma \ref{trivial-lemma-2} gives us also a necessary and sufficient condition for equality in the concavity of the capacity functional.
\section{Conclusion}
We have described the structure of optimal input covariance matrices using the symmetries of the channel matrix $H$ at our disposal. Those symmetries are encoded in the right commutant $\mathcal{C}_{\Sigma}$. If $\mathcal{C}_{\Sigma}\neq\cc \idn_{N}$ the original optimization problem reduces to independent optimization problems coupled only over the trace constraint of Corollary \ref{Cor1}.
\section*{Acknowledgment} We would like to thank Eduard Jorswieck for helpful discussions on this topic . This research was supported by the DFG via projects Bj 57/1-1 ``Entropie und Kodierung gro\ss er Quanten-Informationssysteme'' and Bo 1734/2-1 ``Optimale Sendestrategien f\"ur MIMO unter partieller Kanalkenntnis''.


\begin{thebibliography}{1}

\bibitem{telatar}
E. Telatar, ``Capacity of multi-antenna Gaussian channels,'' \emph{Bell Labs J.} vol. 10, no. 6, 1999

\bibitem{foschini}
G.J. Foschini and M.J. Gans, ``On limits of wireless communications in a fading environment when using multiple antennas,'' \emph{Wireless Pers. Commun.} vol. 6, pp. 311-335, 1998

\bibitem{chuah}
C.N. Chuah, D.N.C. Tse, J.M. Kahn, ``Capacity scaling in MIMO wireless systems under correlated fading,'' \emph{IEEE Trans. Inform. Theory} vol. 48, pp. 637-650, 2002

\bibitem{shiu}
D. Shiu, G.J. Foschini, M.J. Gans, J.M. Kahn, ``Fading correlation and its effect on the capacity of multi-element antenna systems,'' \emph{IEEE Trans. Commun.} pp. 502-513, 2000

\bibitem{simon}
S.H. Simon, A.L. Moustakas, ``Optimizing MIMO antenna systems with channel covariance feedback,'' \emph{IEEE J. Select. Areas Commun.} vol. 21, pp. 406-417, 2003

\bibitem{jafar}
S. Jafar, A. Goldsmith, ``Transmitter optimization and optimality of beamforming for multiple antenna systems,'' \emph{IEEE Trans. Wireless Commun.} vol.3, pp. 1165-1175, 2004

\bibitem{goldsmith}
S.A. Jafar, S. Wishwanath, A. Goldsmith, ``Channel Capacity and Beamforming for Multiple Transmit and Receive Antennas with Covariance Feedback,'' \emph{Proceedings of ICC 2001} 2001

\bibitem{eduard-holger}
E.A. Jorswieck, H. Boche, ``Channel Capacity and Capacity-Range of Beamforming in MIMO Wireless Systems under Correlated Fading with Covariance Feedback,'' \emph{IEEE Trans. Wireless Commun.} Vol. 3, No. 5, 2004


\bibitem{takesaki}
M. Takesaki, ``Theory of Operator Algebras I,'' Encyclopedia of Mathematical Sciences, Springer, Berlin 2000

\bibitem{gurvits}
L. Gurvits, H. Barnum, ``Largest separable balls around maximally mixed bipartite quantum state,'' \emph{Phys. Rev. A} 66, 062311, 2002

\bibitem{werner}
R.F. Werner, ``Quantum states with Einsten-Podolski-Rosen correlations admitting a hidden-variable model,'' \emph{Phys. Rev. A} 40, 4277, 1989

\bibitem{caire}
G. Caire, S. Shamai, ``On the Capacity of Some Channels with Channel State Information,'' \emph{IEEE Trans. Inform. Theory} vol. 45, no. 6, pp. 2007-2019, 1999

\bibitem{horn-johnson}
R.A. Horn, C.R. Johnson, ``Matrix Analysis,'' Cambridge University Press, Cambridge 1999

\bibitem{verdu}
A.M. Tulino, A. Lozano, S. Verd\'u, ``Capacity-Achieving Input Covariance for Single-User Multi-Antenna Channels,'' \emph{IEEE Trans. Wireless Commun.} vol. 5.no. 3 662-671, 2006
\end{thebibliography}
\end{document}